\documentclass[lettersize, journal, doublecolumn]{IEEEtran}
\usepackage[english]{babel}

\usepackage{subcaption}
\DeclareCaptionLabelSeparator{periodspace}{.\quad}
\usepackage[]{footmisc}
\usepackage{cite}
\usepackage{amsmath,amssymb,amsfonts}
\usepackage{algorithmic}
\usepackage{graphicx}
\usepackage{textcomp}
\usepackage{xcolor}
\usepackage{float}
\usepackage{url}
\usepackage{rotating}
\usepackage{gensymb}
\PassOptionsToPackage{hyphens}{url}\usepackage{hyperref}

\def\BibTeX{{\rm B\kern-.05em{\sc i\kern-.025em b}\kern-.08em
    T\kern-.1667em\lower.7ex\hbox{E}\kern-.125emX}}

\makeatletter
\newcommand{\linebreakand}{
  \end{@IEEEauthorhalign}
  \hfill\mbox{}\par
  \mbox{}\hfill\begin{@IEEEauthorhalign}
}
\makeatother

\title{On the Placement and Sustainability of Drone FSO Backhaul Relays}

\author{\IEEEauthorblockN{Salim Janji},~\IEEEmembership{\textit{Graduate} Student Member,~IEEE,
}
\IEEEauthorblockN{Adam Samorzewski},~\IEEEmembership{\textit{Graduate} Student Member,~IEEE,
}
\IEEEauthorblockN{Małgorzata Wasilewska},~\IEEEmembership{\textit{Graduate} Student Member,~IEEE,
}and
\IEEEauthorblockN{Adrian Kliks},~\IEEEmembership{\textit{Senior} Member,~IEEE
}
\thanks{The work has been realized with the statutory funds from Ministerstwo Edukacji i Nauki (MEIN), projects no. 0312/SBAD/8161 and no.0312/SBAD/8159.
Copyright © 2021 IEEE. Personal use is permitted. For any other
purposes, permission must be obtained from the IEEE by emailing pubs-
permissions@ieee.org. This is the author’s version of an article that has
been published in IEEE Wireless Communications Letters on May 27, 2022. Changes were made to this version by the publisher prior to publication, the final version of record is
available at: http://dx.doi.org/10.1109/LWC.2022.3178546.
To cite the paper use: S. Janji, A. Samorzewski, M. Wasilewska and A. Kliks, "On the Placement and Sustainability of Drone FSO Backhaul Relays," in IEEE Wireless Communications Letters, doi: 10.1109/LWC.2022.3178546, or visit
https://ieeexplore.ieee.org/document/9783028}}

\begin{document}
\maketitle

\begin{abstract}
% Free-space optical (FSO) communication links provide reduced interference and larger bandwidth relative to radio frequency (RF) links and they are increasingly considered as a solution for the backhaul connectivity of the emerging networks that leverage the use of small cells (SCs). These SCs can be mounted on unmanned aerial vehicles (UAVs) to make drone base stations (DBSs) that can be dynamically localized to maximize the system performance. Additionally, a UAV with an FSO apparatus can serve as a backhaul relay node. In this paper, we consider how such drone relay stations (DRSs) can be deployed in a high-rise urban area in order to provide FSO line-of-sight (LOS) links that are unobstructed by obstacles (i.e. buildings). Furthermore, since UAVs have limited on-board energy, in our solution we consider the case where solar panels are mounted on DRSs such that placing the DRS in a sunny location is prioritized.
We consider free-space optical (FSO) communication links for the backhaul connectivity of small cells (SCs) where a UAV with an FSO apparatus can serve as a backhaul relay node. We demonstrate how such drone relay stations (DRSs) can be deployed in a high-rise urban area in order to provide FSO line-of-sight (LOS) links that are unobstructed by buildings. Also, in our solution we consider the case where solar panels are mounted on DRSs such that placing the DRS in a sunny location is prioritized, and we show the gain in terms of number of required trips to recharge the UAV.
\end{abstract}

% \linespread{2}
\section{Introduction and Related Work}
Emerging cellular networks increasingly make use of small cells (SCs) as part of the strategy of densifying and shrinking cells in order to fulfill the high data rate and low latency requirements \cite{what5g}. Furthermore, moving small cells such as those mounted on UAVs have received increased research as summarized by the surveys in \cite{survey1} and \cite{survey2}. Drone base stations (DBSs) can adapt their locations according to the changes in users' locations such that path loss and interference are reduced, and they prove to be a cost-effective replacement to fixed small cells structures. However, SCs require a robust backhaul connectivity during their operations \cite{backhaul}, and providing backhaul connectivity to moving SCs is a technical challenge. In that context, free-space optical (FSO) communication links are found to provide superior performance relatively to radio frequency (RF) links because they achieve higher data rates, smaller antenna sizes leading to lower weight and volume, and increased security \cite{fsoVsRf}. Deploying FSO transceivers on UAVs has received attention in recent years.  The authors in \cite{fsoA2Gmodel} present a statistical channel model for FSO communication between a static transceiver and another one mounted on an UAV. In \cite{fsoAlignUav}, the authors analyzed air-to-air (A2A) FSO communications using multirotors and showed that commercial multirotor drones can sustain an A2A FSO link despite the alignment difficulties according to the model they introduced. The authors in \cite{fsoUavBackhaulFronthaul} consider a set of DBSs with FSO transceivers whereas an FSO backhaul is provided to each DBS through FSO A2A links. They show that utilizing FSO for both backhaul and fronthaul substantially improves the system throughput. In \cite{ergodicRate}, the authors were able to derive an approximation for the ergodic rate of an FSO link between an UAV serving as a relay and a ground station. Then, by combining it with an approximation for the ergodic sum rate of users served by the UAV through RF they obtained the ergodic rate of the relay. Similarly, UAVs serve as buffer-enabled FSO relays between stationary nodes in \cite{fsoBufferRelayUav} enhancing the outage probability. However, the problem of placing antennas in locations such that a line-of-sight (LOS) link can be formed with the next hop without getting blocked by an obstacle is not considered since antennas are assumed to exist at a sufficiently large altitude whereas in this work, we consider that the obstacles are high enough to obstruct paths (e.g., high-rise urban area with skyscrapers). DRSs can also be placed at low altitudes in cases where they also establish RF links with ground nodes to limit interference to other nodes \cite{interferenceAltitude}.

Moreover, UAVs flight duration is bounded by the on-board battery lifetime. To improve the energy performance, solutions can optimize for energy efficiency such as in \cite{energyEfficientRef} where the authors derived a model for the propulsion energy consumption taking the trajectory into consideration, and then they utilized it to optimize the trajectory of a UAV communicating with a ground terminal such that the energy efficiency is maximized. Alternatively, to cope with the limited on-board energy that DBSs have, the authors in \cite{solarPoweredUav} demonstrated a UAV prototype with solar panels. Also, the authors in \cite{Sekander2021} study the use of renewable energy sources by UAVs and formulate a signal-to-noise ratio (SNR) outage minimization problem which optimizes the transmit power and flight time of the UAV. Similarly in \cite{uavSolarOptimization}, the authors formulate a joint trajectory and resource allocation optimization problem in which a UAV can increase its altitude in the presence of clouds in order to harvest more energy. An algorithm is proposed to determine the UAV's trajectory in each time slot which takes into consideration the aerodynamic power consumption, on-board energy, and users' quality of service (QoS) requirements. However, in this work we consider the presence of high buildings that can obstruct both the light from the sun and the FSO link, and therefore the placement of DRSs should avoid both types of obstructions.

% Moreover, UAVs flight duration is bounded by the on-board battery lifetime. To cope with the limited on-board energy that DBSs have, the authors in \cite{Sekander2021} study the use of renewable energy sources by UAVs and formulate an signal-to-noise ratio (SNR) outage minimization problem which optimizes the transmit power and flight time of the UAV. Similarly in \cite{uavSolarOptimization}, the authors formulate a joint trajectory and resource allocation optimization problem in which a UAV can increase its altitude in the presence of clouds in order to harvest more energy. An algorithm is proposed to tackle the problem which takes into consideration the aerodynamic power consumption, on-board energy, and users quality of service (QoS) requirements.

In this paper, we consider the problem of placing solar-powered DRSs to ensure backhaul connectivity to a given point between obstacles. First, we propose the use of Lee's visibility graph algorithm \cite{leeAlgorithm} to find the locations of DRSs such that the LOS visibility is ensured between successive hops, and then we modify the algorithm so that placing DRSs in sunny locations is more preferable than in shadowed spots. We show that for the given environment the algorithm is able to place DRSs in sunny spots during all daytime hours. Furthermore, by adopting the energy models mentioned in literature, we show the gain in performance when using solar panels in terms of required number of recharge trips needed per day to sustain backhaul connectivity to a given destination.

The rest of the paper is organized as follows. Section \ref{section:systemModel} describes the considered scenario along with the power consumption model and energy harvesting and storage systems. Section \ref{section:nonsolarPlacement} presents the solution for finding DRSs locations after constructing the visibility graph. In Section \ref{section:solarPlacement} we extend the mechanism such that we also find sunny locations and prioritize choosing them in the shortest path algorithm. Furthermore, Section \ref{section:simulation} presents the simulations results, and Section \ref{section:conclusion} concludes the paper.

\section{System Model}\label{section:systemModel}

\subsection{Environment Description}

% In our work, we consider a cellular network placed in a high-rise urban  area. The wireless system consists of $N$ nodes of UAV type and a single Macrocell Base Station (MBS). The UAVs, scattered all over the area, are responsible for handling the users, and they are powered by photovoltaic (PV) panels mounted on their top covers. On the other hand, the MBS is equipped with on-grid power supply. This node's major objective is to oversee UAVs' behavior, i.e., to coordinate the altitude and target locations of the hovering, and transmit power levels of UAVs' antennas as well. Moreover, the MBS acts as a gateway for backhaul links between access points (UAVs) and the core network of the system. These connections between nodes were implemented according to Free Space Optics (FSO) technology. Thus, there is a need to keep each access node in line-of-sight (LOS) with the gateway (MBS) or at least another UAV in such a way to provide a stable data transfer to this gateway through indirect FSO links. In the considered area, mobile users are deployed randomly and uniformly, and they can change their positions in accordance with the mobility model described in \cite{Baumann2007}. Fig.~\ref{fig_system} presents a simple scheme of the system scenario assumed in our analysis.

\begin{figure}[!t]
\centering
\centerline{\includegraphics[width=0.47\textwidth, angle=0, trim={1cm 0.5cm 1.5cm 1cm},clip]{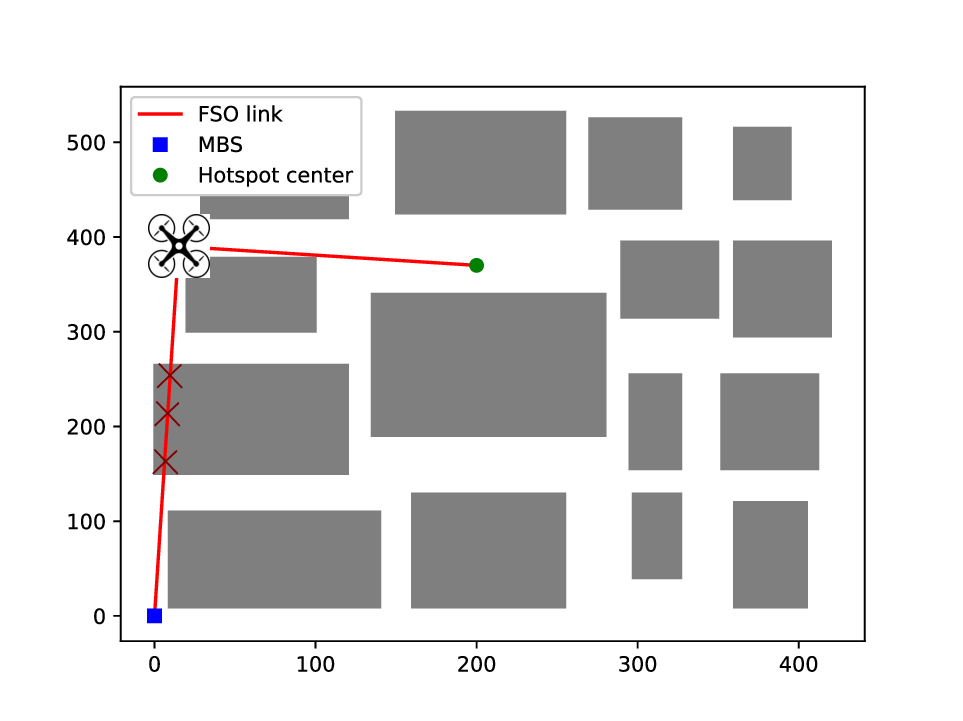}}
\caption{High-rise buildings with MBS and hotspot destination.}
\label{buldings_scenario}
\end{figure}

We consider a high-rise urban area with a macrocell base station (MBS) located at $L_M=(x_M, y_M)$ and a known hotspot location, $L_H=(x_H, y_H)$, as indicated in Fig.~\ref{buldings_scenario}. Hotspots locations can be calculated from users locations that can be obtained using UE localisation schemes such as observed time difference of arrival (OTDoA) \cite{observedTimeDifference} or self-reported GPS coordinates. Hotpsots can also be predicted using historical records and deep learning as demonstrated in \cite{DeepHotspot}. To improve the spectral efficiency and increase total system throughput, a dedicated DBS is deployed in the area of hotspot location \cite{twoTierDeployment}. The deployed DBS is backhaul-connected to the MBS through FSO links provided by a set of DRSs. The channels established by these links can adopt similar assumptions to those presented in \cite{ergodicRate}. We omit them here since this work focuses solely on avoiding obstacles\footnote{Optimizing transmission parameters for reliability and throughput is left for future work.}. Therefore, successive hops should not have any obstacles blocking the FSO path between them. Fig.~\ref{buldings_scenario} shows the case where the link between the MBS and the DRS is blocked by a building, whereas the link between the DRS and hotpost is visible. Furthermore, we assume that the DRSs are equipped with solar panels that provide energy to the on-board batteries whenever the DRSs is in a sunny location. The energy consumption and regeneration models are introduced below.

% \begin{figure}[ht]
% \centerline{\includegraphics[width=0.5\textwidth, angle=0]{pictures/icc-system-scheme-v7.eps}}
% \caption{Considered scenario of the system}
% \label{fig_system}
% \end{figure}

\subsection{Power Consumption Model}

Following \cite{Sekander2021}, for each DRS $n$, $\left(1,\ldots,N\right)$, the aerodynamic energy consumed over observation time $t_{\mathrm{o}}$, is described as $ E_{\mathrm{UAV},n} = P_{\mathrm{hov},n} \cdot t_{\mathrm{o}}$ where $P_{\mathrm{hov},n}$ is the power consumption for hovering determined as:
\begin{align}
    P_{\mathrm{hov},n} = \sqrt{\frac{\left(m_{n}g\right)^{3}}{2\pi r_{\mathrm{p},n}^{2}l_{\mathrm{p},n}\rho}},
    \label{eqUAVflight}
\end{align} 
where $m_{n}$ is the mass of the drone (in kg), and $g$ denotes earth gravity (in $\frac{m}{s^2}$). The expression $2\pi r_{\mathrm{p},n}^{2}l_{\mathrm{p},n}$ describes the total area of the UAV's propellers (in $m^2$), where $l_{\mathrm{p},n}$ is their number and $r_{\mathrm{p},n}$ is the radius of a single propeller, given in meters, and $\rho$ determines the air density (in $\frac{kg}{m^3}$).

Next, the energy consumed to sustain an FSO link is described by $ E_{\mathrm{back},n}\left(t_{\mathrm{o}}\right) =P_{\mathrm{back},n} \cdot t_{\mathrm{o}}$,
% \begin{align}
%     E_{\mathrm{back},n}\left(t_{\mathrm{o}}\right) = l_{\mathrm{c},n} \cdot P_{\mathrm{back}} \cdot t_{\mathrm{o}},
%     \label{eqUAVback}
% \end{align}
where $E_{\mathrm{back},n}$ denotes the energy quantity in Wh spent to sustain the FSO backhaul link, through which the DRS relays the data to other nodes (including the MBS). $P_{\mathrm{back},n}$ defines the power required for the FSO link and is assumed to be constant.
%Next, $l_{\mathrm{c},n}$ is the number of active FSO connections for a particular UAV.

\subsection{Energy Harvesting Model}

To evaluate the performance of the photovoltaic (PV) panels supporting drones, the following energy acquisition model is applied: $E_{\mathrm{PV},n}\left(t_{\mathrm{o}}\right) = P_{\mathrm{PV},n} \cdot t_{\mathrm{o}}$,
% \begin{align}
%     E_{\mathrm{PV},n}\left(t_{\mathrm{o}}\right) = P_{\mathrm{PV},n} \cdot t_{\mathrm{o}},
%     \label{eqPVenergy}
% \end{align}
where $P_{\mathrm{PV},n}$ describes the power generated by a single solar panel mounted on the $n$-th DRS's top cover. Assuming that the inclination angle of the solar collector’s surface is equal to $0\degree$, then as in~\cite{Dimitriou2020}
\begin{align}
    P_{\mathrm{PV},n} = T_{\mathrm{atm}} \cdot I_{\mathrm{0}} \cdot \delta \cdot \eta_{\mathrm{PV},n} \cdot S_{\mathrm{PV},n} \cdot sin\left(\alpha\right),
    \label{eqPVprod}
\end{align} 
where $\alpha$ is the solar altitude angle given in degrees, and $\eta_{\mathrm{PV},n}$ and $S_{\mathrm{PV},n}$ are the PV panel's efficiency factor and surface in square meters, respectively. Next, $T_{\mathrm{atm}} I_{\mathrm{0}}$ is the power of solar irradiation per unit area, where $I_{0}$ denotes the mean solar irradiation outside the atmosphere, both specified in $\mathrm{W/m^2}$, and $T_{\mathrm{atm}}$ specifies the atmospheric transmittance. Finally, to consider the impact of clouds on the acquired solar irradiation (and therefore on the amount of produced energy), the scaling parameter $\delta$ is used, which is a pseudorandom number generated from an uniform distribution between $0.8$ and $1$ that reduces the harvested solar energy.

\subsection{Energy Storage System}

% In order to prevent our system against unwanted waste of available energy resources, each UAV access point was equipped with a storage system in the form of a single accumulator. This approach would ensure an additional energy source for the situation, when the energy demand provoked by the system is higher than the generated amount of it by PV panels (e.g., when the user traffic is too high or when the irradiance of the sun is too weak). 

Assuming perfect charging efficiency, the energy storage in observation time $t_{\mathrm{o}}=t_{s} - t_{s-1}$ is formulated as \cite{Jahid2018} as:
\begin{align}
    &E_{\mathrm{batt},n}\left(t_{s}\right) = \\ &E_{\mathrm{batt},n}\left(t_{s-1}\right) + E_{\mathrm{PV},n}\left(t_\mathrm{o}\right) - E_{\mathrm{UAV},n}\left(t_{\mathrm{o}}\right) - E_{\mathrm{back},n}\left(t_{\mathrm{o}}\right), \nonumber
    \label{eqBattery}
\end{align} 
where $E_{\mathrm{batt},n}\left(t_{s}\right)$ is the energy accumulated inside the battery of the $n$-th drone at $\left(t_{s}\right)$ time stamp. 

% In the considered RES energy generation process, we assumed the impact of shadows cast by the buildings on the energy amounts produced by UAVs' PV panels. Thus, depending on the actually simulated time of the day, the shadow distribution map of the examined area changes in accordance with the sun's elevation (altitude) and azimuth angle, as defined in \cite{Itaca}. 
% The latter can be evaluated based on the following formula \cite{Itaca}:
% \begin{align}
%     sin\left(\gamma\right) = \frac{sin\left(\theta\right)cos\left(\phi\right)}{cos\left(\alpha\right)}
% \end{align}
% where $\gamma$, $\theta$, $\phi$ and $\alpha$ are the angles of the sun's azimuth, hour, declination, and altitude, respectively. All these parameters are given in degrees. 

\section{FSO Backhaul Shortest Path}\label{section:nonsolarPlacement}
\subsection{Problem Formulation}
Let the hotspot location be given by $L_H=(x_H, y_H)$, and the start location relating to the MBS as $L_M=(x_M, y_M)$. Then, the goal is to find a communication path from $L_M$ to $L_H$ that does not pass through any obstacle from the set of buildings shown in Fig.~\ref{buldings_scenario}. Considering that the shortest path between obstacles is a set of straight lines that connects the source to destination via a possibly empty sequence of vertices of obstacles \cite{perezAlgorithm}, our goal can be formulated as a visibility graph problem. In this case, an input graph $G$ describes the set of geometric basic objects (e.g., convex polygons defined as the set of vertices and edges connecting them) and the resulting solution of the problem is a visibility graph $G_v$ which contains all possible edges between any two vertices in $G$ such that these two vertices are said to be \textit{unobstructed} by any obstacle (i.e., having an obstacle-free LOS path between them). Then, the final solution can be obtained by applying a shortest path algorithm (e.g. Dijkstra's) to resulting graph to select the locations in which DRSs should be deployed to serve as backhaul nodes. In this paper we utilize Lee's algorithm presented in \cite{leeAlgorithm}, taking advantage from its simplicity.

\subsection{Proposed DRS Placement Algorithm}\label{drsNoSun}
Given a map of the obstacles, we extract the graph $G(V, E)$ containing all the vertices $V$ and edges $E$ that define the buildings (i.e. building corners and building edges in the top view, respectively). We add to the set of vertices, $V$, the two points defining the start and end locations given by $L_M$ and $L_H$, respectively, so that their visibility tree is also computed. Then we implement Lee's algorithm for each $v_i \in V$ to obtain the complete visibility graph $G_v$ as follows~\cite{coleman2012lee}:
\begin{enumerate}
    \item Initialize a horizontal scan line, $\vec{s}$, starting from the center vertex $c=v_i$ and pointing rightwards.
    \item Sort all other vertices $\{v_j \in V - v_i\}$ in ascending order according to the angle $\theta_j$ between $\vec{s}$ and $\vec{cv}_j$ (i.e. by rotating $\vec{s}$ counter-clockwise) inside a data structure $A$.
    \item In another data structure, $E_s$, sort all the edges that are intersecting with $\vec{s}$ in ascending order according to their distance from $c$, along with the distance magnitude. Clearly, the first edge is the only one that $c$ can see.
    \item Rotate $\vec{s}$ to every vertex $v_j$ in order of $\theta_j$ in $A$; at every vertex, check any edge that $v_j$ is a part of, and either add or remove this edge from $E_s$ appropriately if $\vec{s}$ would intersect this edge or not at the next rotation step.
    \item At every rotation step check if the distance to $v_j$ is less than the distance to the closest edge in $E_s$; if so, then $v_j$ is visible and it is added to the visibility tree of $v_i$, otherwise not. 
\end{enumerate}

After obtaining $G_v (V_g, E_g)$, each edge $E_k=\{v_i, v_j\}$ is assigned a cost given by
\begin{equation}\label{eqn:cost}
    D = \frac{d(v_i, v_j)}{d_{\max}} + 1,
\end{equation}
whereas $d(v_i, v_j)$ is the Euclidean distance between the two endpoints, and $d_{\max}$ is a normalizing factor which limits the distance cost to $[0, 1]$ and can be set to the maximum distance between any two points in the considered area. The reason for defining the cost in such a way is that we want to obtain the shortest path in terms of hops primarily and then considering the link distance as a secondary objective. The additional value of $1$ causes that choosing an edge increases the number of hops and therefore the number of DRSs. Finally, we can apply Dijkstra's algorithm on $G_v$ with edges costs defined as in \eqref{eqn:cost} to find the shortest communications path.

\section{Proposed Solar Powered DRS Placement}\label{section:solarPlacement}
To improve energy efficiency, in this paper we consider solar-powered DRSs and target the problem of their placement in a sunny environment with shadows resulting from buildings. The goal is to leverage the previously introduced energy generation system in order to prolong the deployment duration of DRSs and reduce the number of trips required by each DRS to return to its deployment base in order to recharge its battery. 

\subsection{Sunny Points Search}\label{subsection:pointSearch}
\begin{figure}[!htb]
\centerline{\includegraphics[angle=0]{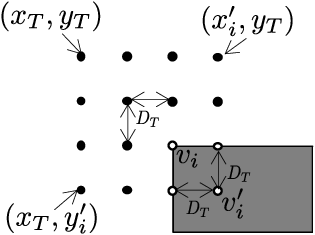}}
\caption{Test points grid for finding sunny spots for $C_T=4$.}
\label{sun_grid}
\end{figure}

Since the DRSs should be placed in sunny locations, the first step is to find the set of possible points at a given hour in a day that are sunny (i.e., not shadowed by any building). A good starting point which guides us towards finding the shortest path is the set of building vertices $V$ which we used in the previous section. Thus, given the sun altitude and azimuth angles, $\alpha$ and $\phi$, and the set of obstacles defined by $G(V, E)$, we are able to test any point of interest if it falls within any shadow caused by a building in the direction $-\phi$ with a length of $\frac{h_\text{B} - h_\text{UAV}}{tan(\alpha)}$ such that $h_\text{B}$ and $h_\text{UAV}$ define the heights of the building and the UAV, respectively. To find the set of sunny points $V_s$ we perform the following steps for every vertex $ v_i=(x_i, y_i) \in V$. Let $v'_i = (x'_i, y'_i)$ be the point in the interior of the polygon of which $v_i$ is a vertex such that $v'_i$ is equidistant from the two edges connected to $v_i$ with a distance of $D_T$ (see  Fig.~\ref{sun_grid}). First, we create a grid of test points around $v_i$ denoted as $V_T = \{(x_j, y_j) \; |\; x_j = x'_i \pm n D_T, \; y_j = y'_i \pm n D_T, \text{ and } n=1,2,..,C_T, \text{ and } (x_j, y_j) \notin \{(x'_i,y'_i), (x'_i,y_i), (x_i,y'_i), (x_i,y_i)\}\}$ whereas we perform subtraction along the x-axis when $x_T < x'_i$, otherwise addition. Likewise along the y-axis. $C_T$ and $D_T$ determine the number of test points and the spacing between them, respectively. An example is shown in Fig.~\ref{sun_grid}. The initial point is invalid because it's within the building, and we discard the points that are on the edges to consider irregularities in the building surface that could obstruct the light (e.g. balconies, advertisements, etc.). These points are then sequentially tested starting with the points closest to the building such that if and when a point is found to be in a sunny spot, then this point is added to $V_s$ and the search stops.

% To find the set of sunny points $V_s$ we perform the following search for every vertex $ v_i=(x_i, y_i) \in V$. First, we create a grid of test points of size $C_T \times C_T - 4$ that defines the points of interest around $v_i$ as shown in Fig.~\ref{sun_grid}. Adjacent test points are separated by a distance of $D_T$ which is defined as the search step size along each axis. Let $v'_i = (x'_i, y'_i)$ be the point in the interior of the polygon of which $v_i$ is a vertex such that $v'_i$ is equidistant from the two edges connected to $v_i$ with a distance of $D_T$. Starting from $v'_i$ and after $C_T \times C_T$ steps we reach $(x_T,y_T)$. Then the set of test points is defined as $V_T = \{P=(x_j, y_j) \; |\; x_j = x'_i \pm n D_T, \; y_j = y'_i \pm n D_T, \text{ and } n=1,2,..,C_T, \text{ and } P\notin \{(x'_i,y'_i), (x'_i,y_i), (x_i,y'_i), (x_i,y_i)\}\}$ whereas we perform subtraction along the x-axis when $x_T < x'_i$, otherwise addition. Likewise along the y-axis. The initial point is invalid because it's within the building, and we discard the points that are on the edges to consider irregularities in the building surface that could obstruct the light (e.g. balconies, advertisements, etc.). These points are then sequentially tested starting with the points closest to the building such that if and when a point is found to be in a sunny spot, then this point is added to $V_s$ and the search stops.

\subsection{DRS Placement in Sunny Spots Algorithm}
Having obtained the set $V_s$ of sunny locations, we can now apply Lee's algorithm (see Subsection \ref{drsNoSun}) however this time with the set of vertices (and points) being defined as $V' = \{V 	\cup V_s \cup \{L_H, L_M\}\}$. This will also result in obtaining the visibility trees for the new points of $V_s$. After obtaining the visibility graph, we again perform Dijkstra’s algorithm with the cost function defined for every edge $E_k=\{v_i, v_j\}$ as follows

\begin{equation}
    D = 
    \begin{cases}
         \frac{d(v_i, v_j)}{d_{\max}} + 1, & \text{if   $v_j \in V_s$},\\
         \frac{d(v_i, v_j)}{d_{\max}} + 100 & \text{otherwise}.
    \end{cases}
    \label{eqn:newCost}
\end{equation}

The cost function defined in \eqref{eqn:newCost} aims to penalize selecting a shadowed spot by increasing the hop penalty to $100$. This results in avoiding any shadowed spots whenever possible.

\section{Simulation and Results}\label{section:simulation}
In simulations we considered a location in Madrid, Spain, for the solar calculations during the  whole day period of 24 hours. We assumed the environment model shown in Fig.~\ref{buldings_scenario} where a backhaul is provided to the drone serving users at the hotspot $L_H$. The simulation parameters are listed in Table \ref{tab:params}.

\begin{table}
\begin{center}
\caption{Simulation Parameters}
\label{tab:params}
\begin{tabular}{|| p{5.75cm} | c ||}
\hline  Parameter & Value\\
\hline
\hline  Longitude & $3.70427144\;\degree \mathrm{W}$\\
\hline  Latitude & $40.41872533\;\degree \mathrm{N}$\\
\hline  UAV hovering height $\left(h_{\text{UAV}}\right)$ & $20\;\mathrm{m}$\\
\hline  Mean solar irradiation outside the atmosphere $\left(I_0\right)$ & $1353\;\mathrm{W/m^2}$ \cite{Dimitriou2020}\\
\hline  Efficiency of used solar cell type $\left(\eta_\mathrm{PV}\right)$ & $0.2$\\
\hline  Area covered by the solar panels $\left(S_{\mathrm{PV}}\right)$ & $1\;\mathrm{m^2}$ \cite{uavSolarOptimization}\\
\hline  UAV's mass $\left(m\right)$ & $4\;\mathrm{kg}$ \cite{uavSolarOptimization}\\
\hline  Radius of UAV's single propeller $\left(r_\mathrm{p}\right)$ & $0.25\;\mathrm{m}$\\
\hline  Number of UAV's propellers $\left(l_\mathrm{p}\right)$ & $4$\\
\hline  UAV's battery capacity & $222\;\mathrm{Wh}$ \cite{uavSolarOptimization}\\
% \hline  UAV's antenna transmit power $\left(P_\mathrm{t}\right)$ & $100\;\mathrm{mW}$\\
\hline  FSO backhaul link's power consumption $\left(P_\mathrm{back}\right)$ & $200\;\mathrm{mW}$ \cite{fsoUavBackhaulFronthaul}\\
\hline  Distance cost normalizing factor $\left(d_{\max}\right)$ & $700\;\mathrm{m}$\\
\hline  Width of test points $\left(C_T\right)$ &$5 $\\
\hline  Spacing between test points $\left(D_T\right)$& $7 \;\mathrm{m}$\\
\hline
\end{tabular}
\end{center}
\end{table}

\subsection{Placement of DRSs}

\begin{figure*}[!htb]
\centering
\begin{subfigure}{0.32\textwidth}
\includegraphics[width=\textwidth, trim={1cm 0.5cm 1.5cm 1cm},clip]{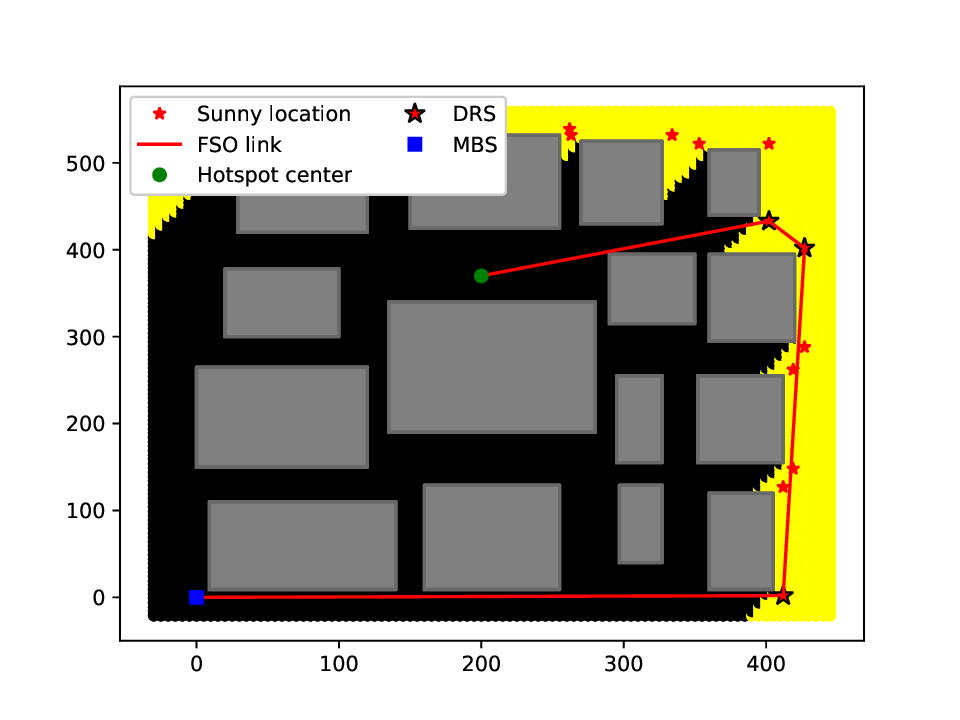}%
\caption{5 am, 3 DRS hops}%
\label{7am}%
\end{subfigure}
\begin{subfigure}{.32\textwidth}
\includegraphics[width=\textwidth, trim={1cm 0.5cm 1.5cm 1cm},clip]{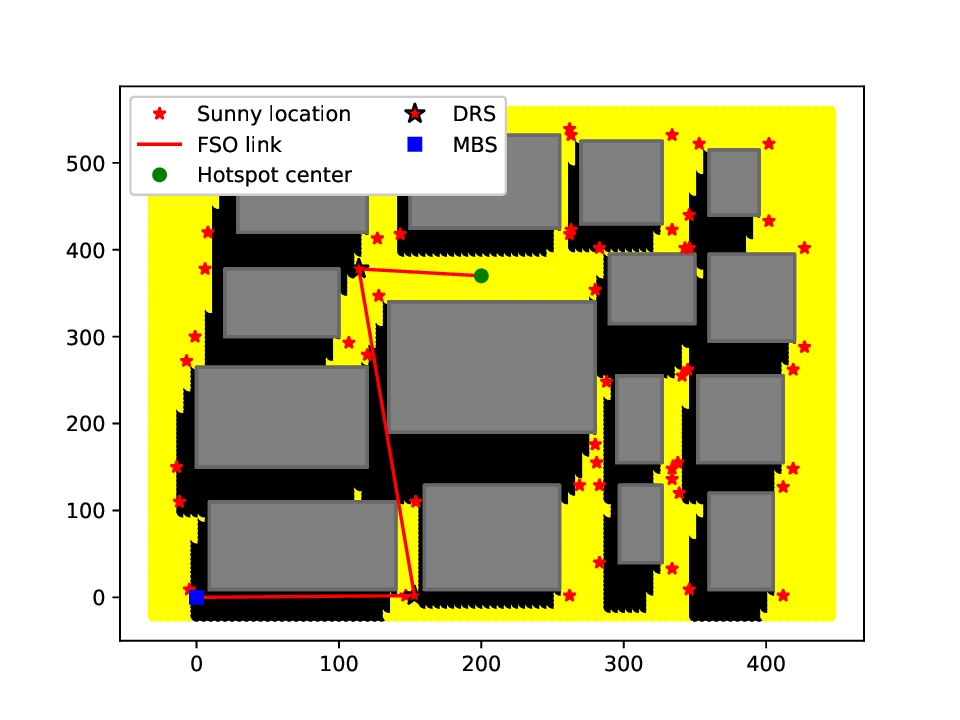}%
\caption{7 am, 2 DRS hops}%
\label{8am}%
\end{subfigure}
\begin{subfigure}{.32\textwidth}
\includegraphics[width=\textwidth, trim={1cm 0.5cm 1.5cm 1cm},clip]{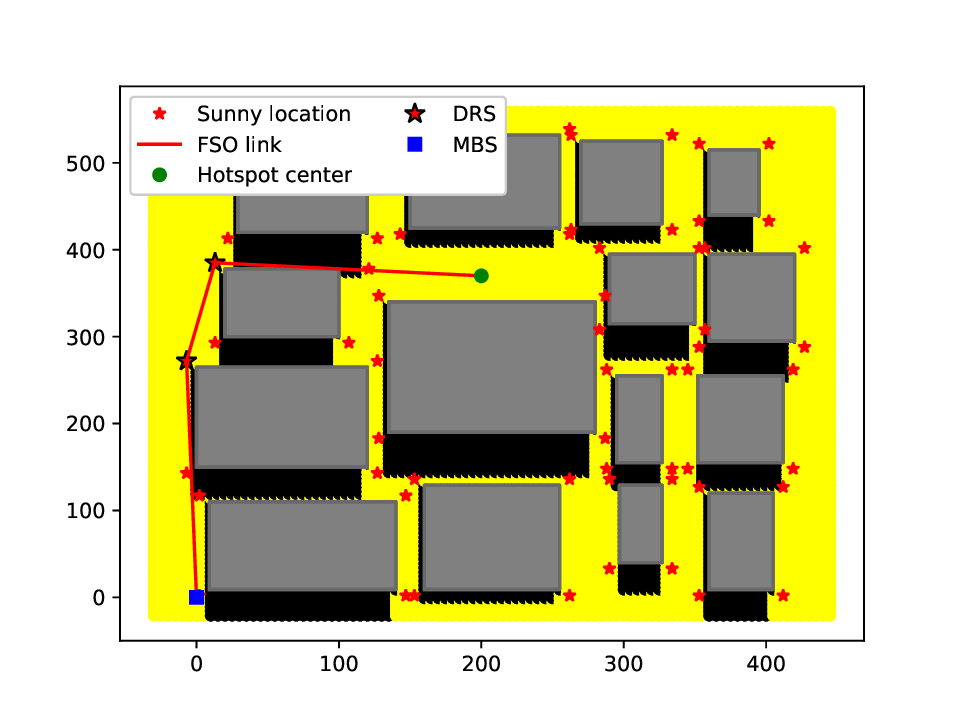}%
\caption{8 am, 2 DRS hops}%
\label{subfigc}%
\end{subfigure}%

\begin{subfigure}{.32\textwidth}
\includegraphics[width=\textwidth, trim={1cm 0.5cm 1.5cm 1cm},clip]{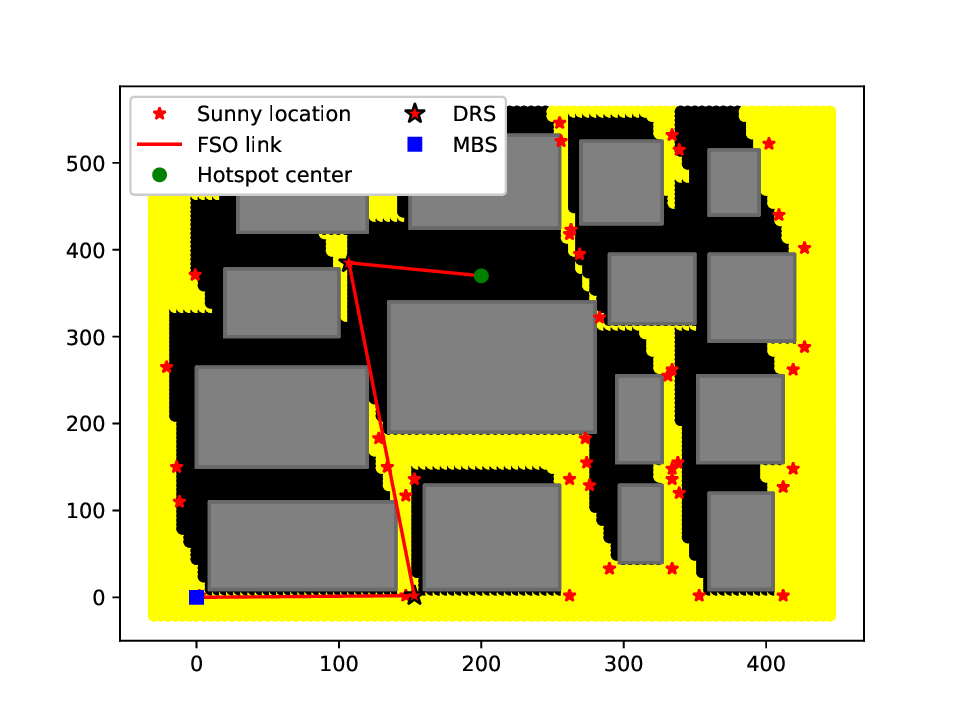}%
\caption{6 pm, 2 DRS hops}%
\label{8pm}%
\end{subfigure}
\begin{subfigure}{.32\textwidth}
\includegraphics[width=\textwidth, trim={1cm 0.5cm 1.5cm 1cm},clip]{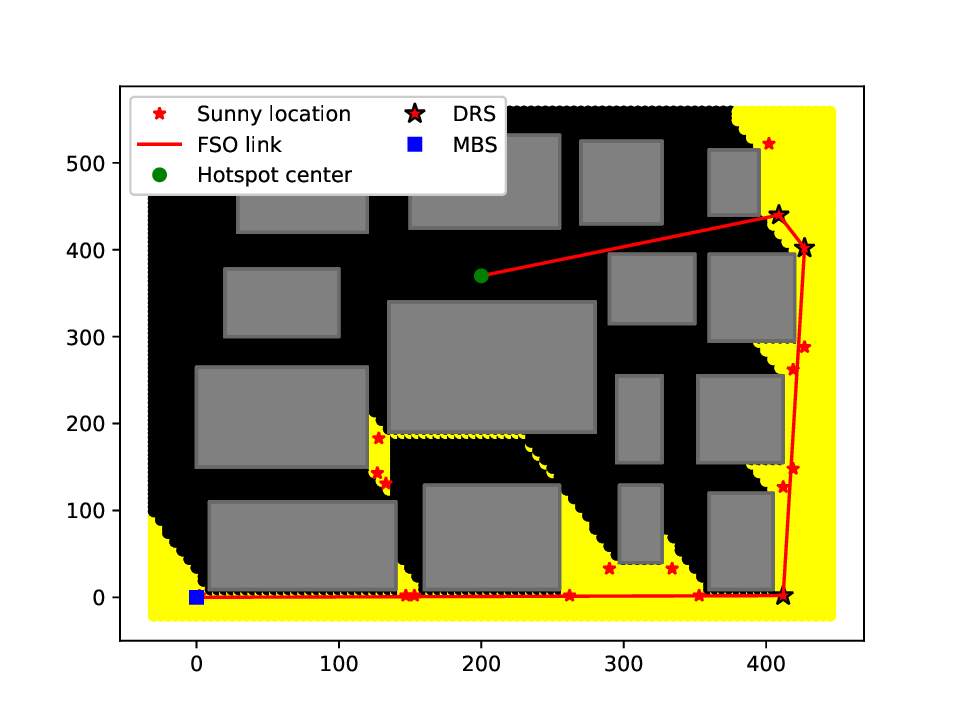}%
\caption{7 pm, 3 DRS hops}%
\label{7pm}%
\end{subfigure}
\begin{subfigure}{.32\textwidth}
\includegraphics[width=\textwidth, trim={1cm 0.5cm 1.5cm 1cm},clip]{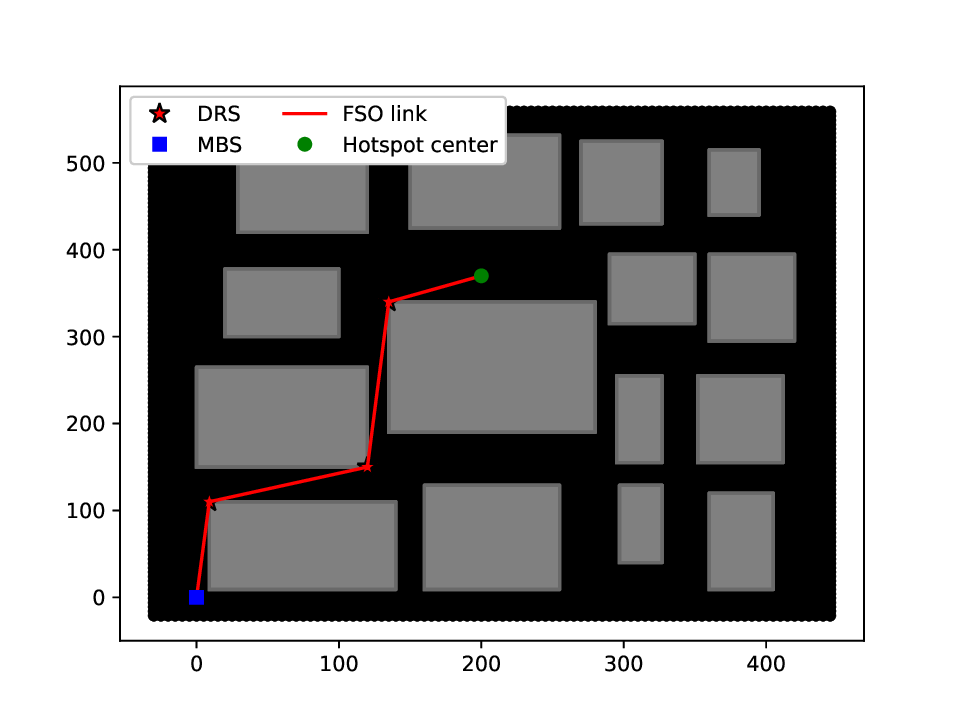}%
\caption{Night, 3 DRS hops}%
\label{night}%
\end{subfigure}
\caption{DRSs placements at different hours of the day.}
\label{drsPlacement}
\end{figure*}

Figure \ref{drsPlacement} shows the placement locations of DRSs along with the shadowed areas for 6 different timings during the day, including the whole night period with no sun. The algorithm is able to find a path of sunny locations at all time instants without selecting any shadowed hop as can be seen. An interesting observation is that the number of hops or required number of DRSs is reduced to two with a total traversed distance of 617 m between 7 am and 7 pm compared with three required hops during the night and a total distance of 490 m. This indicates that the criteria of finding the shortest path using obstacles vertices might not be the best choice whereas such policy does not always yield the least hops path, however, it does indeed result in the shortest path. A different point selection mechanism which would focus on reducing the number of hops and increasing the efficiency of the search relatively to the mechanism introduced in this work can be the target of future work with the help pre-calculated sun maps.

\subsection{Sustainability}
\begin{figure*}[!htb]
\centering
\begin{subfigure}{0.49\textwidth}
\centering
\centerline{\includegraphics[width=1\textwidth, trim={0.2cm 0.2cm 0.2cm 0cm},clip]{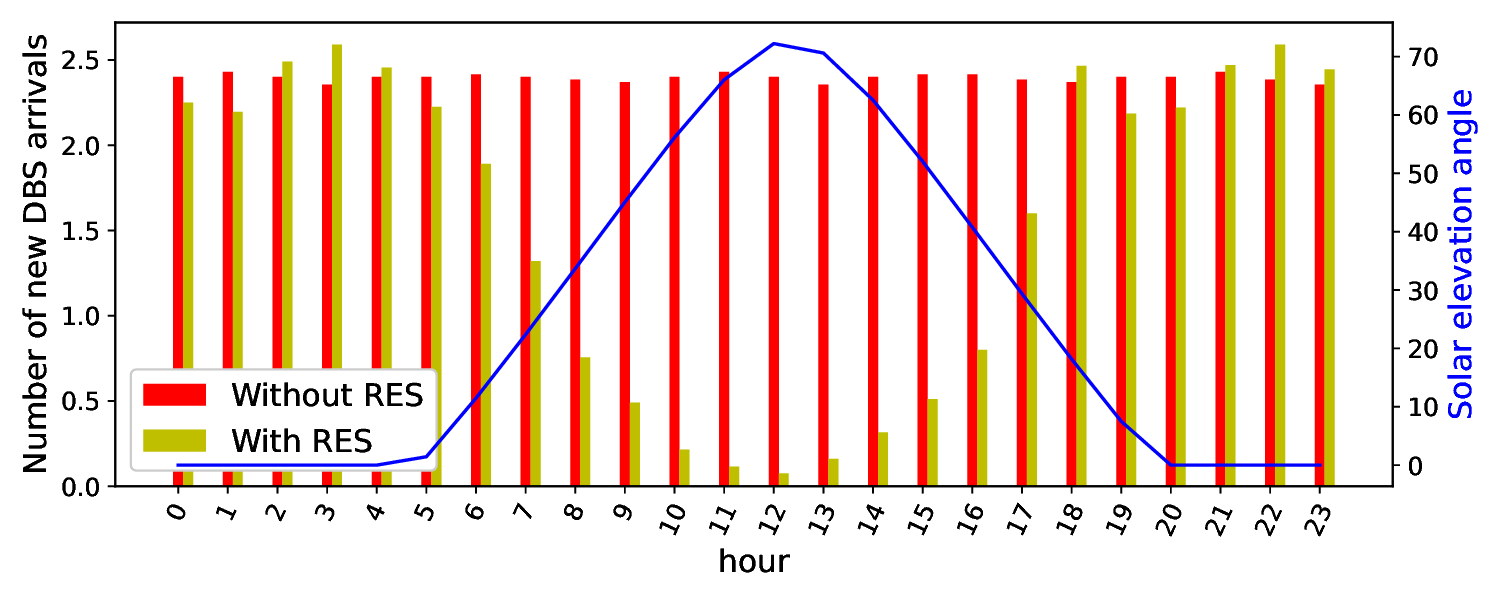}}
\caption{Average number of new DRS arrivals per hour.}
\label{n_arrivals}
\end{subfigure}
\begin{subfigure}{0.49\textwidth}
\centering
\centerline{\includegraphics[width=1\textwidth, trim={0.2cm 0.2cm 0.2cm 0cm},clip]{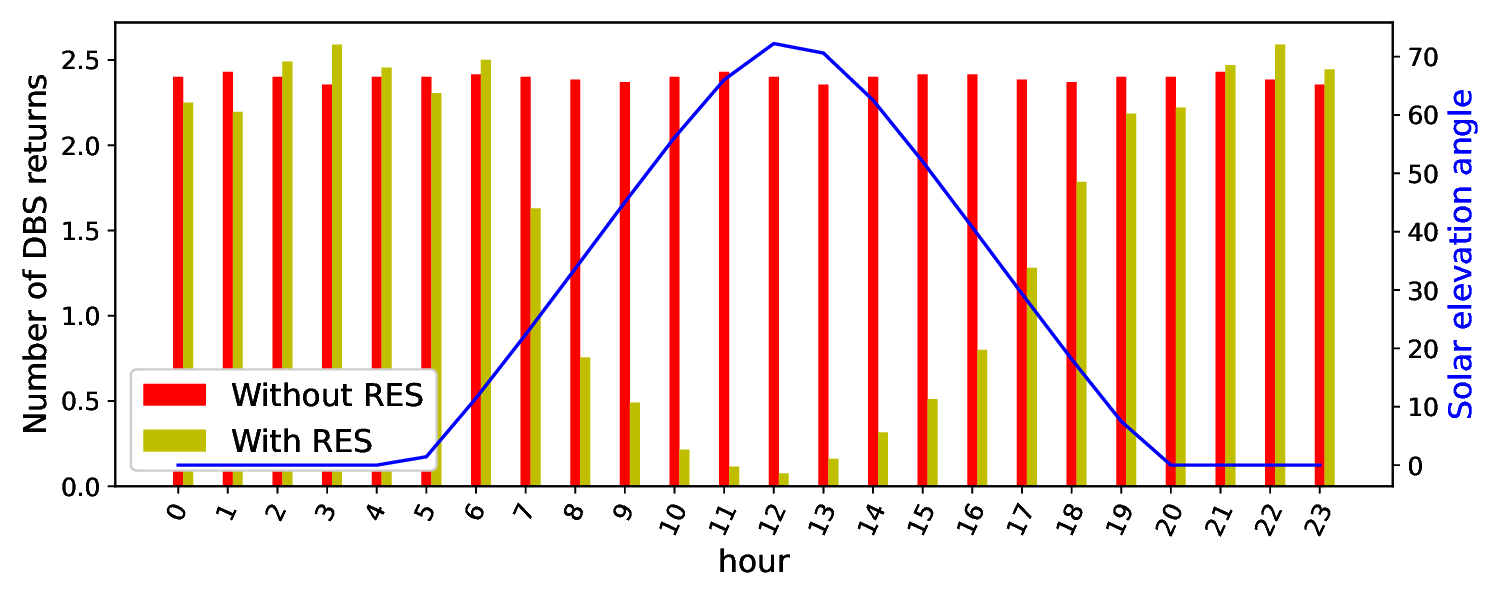}}
\caption{Average number of new DRS returns per hour.}
\label{n_returns}
\end{subfigure}
\caption{Required number of DRS trips back and forth to the charging location.}
\end{figure*}

We can observe the number of new DRS arrivals from deployment center per hour in Figure \ref{n_arrivals}, and the number of DRS return flights per hour in Figure \ref{n_returns}. Arrivals and returns can occur two cases. Either to recharge, or because a new DRS is needed or became redundant (notice the extra arrival at 18 in Figure \ref{n_arrivals} because a new drone is needed as shown in Figure \ref{7pm}). Both figures plot the results with and without RESs. Clearly both counts are greatly reduced during the hours where the sun elevation angle is above zero in the case where solar panels are employed. In total, over the duration of 24 hours, 115 trips of DRSs are required to provide a backhaul to the hotspot location if RESs are not employed. In the case when RESs are used, around 74 trips are needed. That is 35\% less required amount of battery recharges which amounts to $(\frac{115}{2} - \frac{75}{2})222 Wh = 4440 Wh$. Also, depending on the distance to the deployment center, the reduction in energy required for flying back can be substantial in scenarios where the DRSs have to travel a considerable distance. 

\section{Conclusions}\label{section:conclusion}
We have considered obstacles in the placement of DRSs utilizing FSO links and solar panels which was not targeted before in literature to our best knowledge. In this context, we presented how visibility graphs and shortest path algorithms can be leveraged to solve this type of problems. Furthermore, for a given scenario, we showed the gain in utilizing solar panels in terms of required number of recharge trips. 
%We have presented our proposed solution to the problem of FSO backhaul provision using a set of DRSs that are deployed with consideration of existing obstacles using a visibility graph. %Furthermore, we considered employing RESs in the form of solar panels installed on DRSs, and extended the algorithm by the sunny points search presented in Subsection \ref{subsection:pointSearch} and showed that given the set of obstacles presented in Figure \ref{buldings_scenario} the algorithm was always able to find sunny locations in which DRSs are placed such that they can make use of the energy harvested from the sun.
% The simulation results showed a substantial decrease in the required number of flights needed to recharge DRSs. Therefore, a considerable decrease in non-renewable energy utilization is observed. The percentage of this decrease is even expected to increase in the practical scenarios where a hotspot location require service only in daytime hours in comparison to the 24-hours results we presented.

\bibliographystyle{ieeetr}
\bibliography{references}

\end{document}